\documentclass[aps,prb,twocolumn,showpacs,preprintnumbers,amsmath,amssymb]{revtex4-1}
\usepackage{amsmath,amssymb,amsfonts}
\usepackage{bm}
\usepackage{graphicx}
\usepackage{times}
\usepackage{color}
\usepackage[colorlinks,bookmarks=false,citecolor=blue,linkcolor=red,urlcolor=blue]{hyperref}

\newcommand{\fref}[1]{Fig.~\ref{#1}}
\newcommand{\eref}[1]{Eq.~(\ref{#1})}

\newcommand{\normwidth}{0.8\columnwidth}

\newcommand{\bbf}{\textrm}

\begin{document}

\title{Competition between reduced delocalization and charge transfer effects for a two-band Hubbard model}
\author{Reza Nourafkan}
\affiliation{Department of Physics, University of Alberta, Edmonton, Alberta, Canada T6G 2G7}
\author{Frank Marsiglio}
\affiliation{Department of Physics, University of Alberta, Edmonton, Alberta, Canada T6G 2G7}

\begin{abstract} 
We use the embedding approach for a dynamical mean-field method to investigate the electronic properties of a semi-infinite two band Hubbard model at half- and quarter-filling. Two effects determine the degree of correlation at the surface: first, there will charge transfer between the surface and the bulk, and, secondly, electrons at the surface are less delocalized due to the reduced coordination number. We determine the result of these two effects and compute the  quasiparticle weight. It is shown that depletion of charge from the surface to the bulk at quarter-filling competes with enhanced correlation effects;
the net result is that at quarter-filling the quasi particle weight at the surface is approximately equal to the bulk quasi particle weight. Only when the charge transfer approaches zero at large interaction strengths does the quasi particle weight at the surface become lower than that in the bulk.
\end{abstract}
\pacs{71.38.-k, 71.30.+h, 73.20.-r, 71.38.Ht}
\maketitle

\section{Introduction}
Theoretical research on strongly correlated electrons in inhomogeneous systems within the framework of a half-filled semi-infinite single-band model has shown that both the reduced coordination number and (probably) reduced hopping elements near the surface 
cause an enhancement in correlation effects at the surface relative to the bulk.\cite{Potthoff99_1, Potthoff99_2, Borghi09, Nourafkan09, Nourafkan10, Liebsch03, Liebsch06} This in turn leads to an exponential decay of the quasi-particle weight $z$,\cite{Borghi09} as one approaches the surface from the bulk. A lower value of  $z$ at the surface means the coherent peak near the Fermi energy has a lower residue at the surface and the incoherent Hubbard band is more pronounced than in the bulk, in agreement with surface sensitive photoemission experiments on transition-metal oxides such as Ca$_{1-x}$Sr$_{x}$VO$_3$ and La$_{1-x}$Ca$_{x}$VO$_3$.\cite{Maiti98, Maiti04, Maiti01, Sekiyama04} 

Recently, we emphasized the role played by charge transfer between the bulk and the surface, when doping a Mott insulator, and concluded that in approaching half-filling, and for large Coulomb interaction, charge accumulates at the surface from the bulk. \cite{Nourafkan11} A larger charge density at the surface enhances correlation effects and causes a reduced quasi-particle weight at the surface. Therefore, at the surface of a strongly correlated electron system described by a single-band Hubbard model, two mechanisms lead to enhanced correlation effects: weaker delocalization 
at the surface, and charge transfer between the bulk and surface layers.
We note that charge transfer has also played an important role in understanding the conducting properties of interfaces between a band and a Mott insulator,\cite{Ohtomo02, Okamoto04_1, Okamoto04_2, Okamoto04_3, Freericks04} and the reader is referred to the references for further information.

Although in a single band Hubbard model these two mechanisms both contribute to increasing correlation effects on the surface, their effects in a two band Hubbard model at quarter filling actually compete with one another. Indeed, due to depletion of charge from the surface to the bulk at quarter filling, the surface layer will have a reduced correlation, while reduced delocalization at the surface leads to an enhancement of correlation. Therefore, the two band Hubbard model at quarter-filling results in a situation where the interplay between these two mechanisms can be investigated. The purpose of this paper is to study this interplay to see if either mechanism dominates under various circumstances.   

The outline of this paper is as follows. In Sec. II we introduce the model Hamiltonian, which is a semi-infinite two-band Hubbard model with layer dependent parameters. After a very brief description of the dynamical mean field theory procedure used in combination with the embedding technique, we present and analyze results in Sec. III. A summary is provided in Sec. IV.

\section{The Model and Method}
Our aim is to model specifically the phenomena occurring at the surface of a strongly correlated electron system that is governed by the two-band Hubbard model. The system is a three-dimensional,
bipartite simple-cubic (sc) lattice with nearest-neighbour hopping only. The lattice is cut along a plane perpendicular to one
of the coordinate axes, e.g. the $z$-axis [sc(001) surface].
For purposes of calculation, the system is considered to be built up of two-dimensional layers parallel to the surface.
Accordingly, the position vector to a particular site in the
semi-infinite lattice is written as ${\bm R}_{site}={\bm r}_{i}+{\bm
R}_{\alpha}$. Here ${\bm R}_{\alpha}$ stands for the coordinate
origin in the layer $\alpha$, and the layer index runs from
$\alpha=1$ for the topmost surface layer to infinity. The vector ${\bm r}_{i}$ is
the position vector with respect to a layer-dependent origin, and
runs over the sites within the layer. Each lattice site is then labelled by
indices $i$ and $\alpha$. In order to keep the problem simple, the interaction parameters are considered layer independent. To simulate different correlation strengths and relaxation processes at the surface we allow for a possible modification of the hopping between the surface and the subsurface layer $t_{i1,j2}=t_{12}\delta_{ij}$. For simplicity we also assume that the hopping elements are the same for the two orbitals.
In this notation, the Hamiltonian reads:
\begin{eqnarray}
H=&-&\sum_{\langle i\alpha , j\beta \rangle m\sigma}t_{i\alpha,j\beta}  d_{i\alpha m\sigma}^{\dagger}d_{j\beta m\sigma} \nonumber \\ &+& U\sum_{i\alpha m} n_{i\alpha m\uparrow}n_{i\alpha m\downarrow} + \sum_{i\alpha m\sigma}V_{i\alpha} n_{i\alpha m\sigma}\nonumber \\ 
&+& \sum_{i\alpha,\sigma \sigma^{\prime}} \sum_{m>m^{\prime}}(U^{\prime}-J\delta_{\sigma \sigma^{\prime}})n_{i\alpha m\sigma}n_{i\alpha m^{\prime}\sigma^{\prime}} \nonumber \\ 
&-&J\sum_{i\alpha \sigma} \sum_{m\ne m^{\prime}} d^{\dagger}_{i\alpha m\sigma}d_{i\alpha m\bar{\sigma}}d^{\dagger}_{i\alpha m^{\prime}\bar{\sigma}}d_{i\alpha m^{\prime}\sigma}\nonumber \\ 
&-&J\sum_{i\alpha \sigma}\sum_{m\ne m^{\prime}} d^{\dagger}_{i\alpha m\sigma}d^{\dagger}_{i\alpha m\bar{\sigma}}d_{i\alpha m^{\prime}\sigma}d_{i\alpha m^{\prime}\bar{\sigma}}.
\label{ham}
\end{eqnarray}
where $d_{i\alpha m \sigma}\left(d^\dagger_{i\alpha m \sigma} \right)$ is the
destruction (creation) operator for electrons with spin
$\sigma$ and orbital index $m$ on site $i$ of the $\alpha$ layer. The orbital's electron density on site $i\alpha$ is denoted $n_{i\alpha m}$, and $t_{i\alpha,j\beta}$ is the hopping
matrix element between two nearest-neighbour sites. $U$ is the intra-orbital Coulomb interaction, $U^{\prime}$ is the inter-orbital Coulomb interaction, 
and $J$ is the Hund coupling. The last line of \eref{ham} shows the pair-hopping terms. We fix the energy scale by setting $t_{\langle
i\alpha,j\beta\rangle}\equiv t = 1$ for $\alpha,\beta \neq 1$ and we adopt the conventional choice of parameters, $U^{\prime}=U-J$, which follows from symmetry considerations and, for definiteness, we set $J=U/4$. The results presented here are also valid for other values of $J$ (provided $U^\prime - J$ remains positive).

\bbf{The potential $V_{i\alpha}$ shows the electrical potential at the site $i\alpha$, which comes from a redistribution of the charge near the surface; this potential obeys the Poisson equation. For a bulk lattice, in a phase with translational invariance, the local occupation is independent of site, $\langle n_{i \alpha\uparrow}\rangle +\langle n_{i \alpha\downarrow}\rangle = \langle n_{i \alpha}\rangle=n$. On the contrary, for a semi-infinite lattice, the different local environment of the surface sites causes the local occupation near the surface to differ from the bulk filling. This charge redistribution on the layers near the surface gives rise to an electrical potential that couples to electrons and modifies the Hamiltonian. Due to two-dimensional translational invariance  parallel to the surafce, the spatially varying potential is a constant for all sites in a layer parallel to the surface, $V_{i\alpha} = V_\alpha$, and should be determined self-consistently. }

Our calculations are based on the embedding approach\cite{Ishida09,Nourafkan10} for dynamical mean field theory (DMFT)\cite{Georges96} for a simple cubic lattice. \bbf{This method is based on partitioning of the layered structure into a surface region which includes the first $N$ layers and an adjacent semi-infinite bulk region (substrate) to which it is coupled. Then the effect of the substrate on the surface region is described by an energy-dependent embedding potential. The electrical potential $V_{\alpha}$ in the surface region obeys the $1D$ Poisson equation, 
whose discretized solution is
\begin{equation}
V_{\alpha}=\lambda \sum_{\gamma = \alpha}^{N} (\gamma - \alpha + 1) (n_{\gamma}-n_{\rm bulk}) + V_{\rm bulk}
\label{Poisson}
\end{equation}
where $n_{\gamma}$ is the electron density in layer $\gamma$, $n_{\rm bulk}$ is the bulk electron density and $\lambda=e^2/\epsilon a$ with $\epsilon$ the background dielectric constant and $a$ the inter-planar lattice constant. This solution is obtained by iterating the discretized version of the Poisson equation from the bulk layers up to the surface. Hereafter we set $V_{\rm bulk}=0$.} 

In this study, the number of surface layers is chosen to be $N=5$ and we tested that this number provides converged results. Our impurity solver is exact diagonalization.\cite{Caffarel94} The reader is referred to the references for details concerning these methods.
 
\section{Results}

DMFT studies of the two-band Hubbard model for the bulk system have shown the existence of a Mott metal-insulator phase transition at any commensurate band filling, such as quarter-filling ($n=1$) and half-filling ($n=2$). \cite{Koga02, Koga07} For $J=0$ the corresponding critical interaction $U_c$ is maximum for half-filling. Including a non-zero value of $J$ has two competing effects, depending on the filling: at half-filling a non-zero $J$ lowers the critical interaction strength $U_c$, while for all other commensurate fillings the $U_c$ is pushed to very high values by $J$. 

\begin{figure}
\begin{center}
\center{\includegraphics[width=\normwidth]{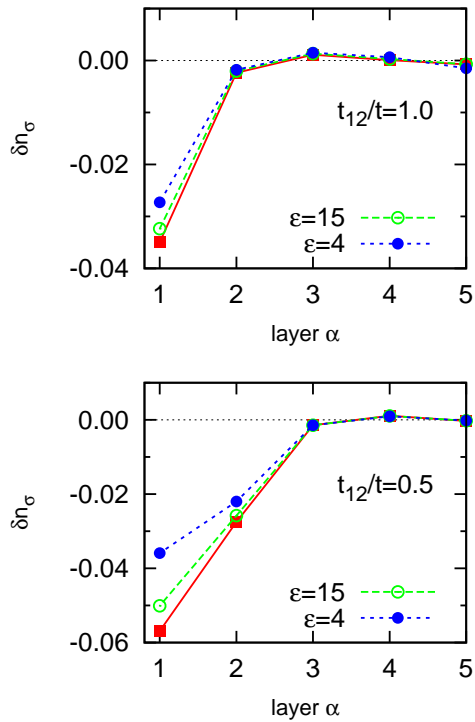}}
\caption{Variation in layer occupation of layers in surface region for $U/D=1/6$ and three values of background dielectric constant. $\alpha = 1$ shows the topmost layer. Top panel: $t_{12}/t=1.0$. Bottom panel: $t_{12}/t=0.5$.}\label{fig:ChargeProfile}
\end{center}
\end{figure}

\bbf{As we mentioned earlier, in a semi-infinite system the physical quantities near surface are layer-dependent. In particular, the surface electron occupation can differ from that of the bulk. We first study this phenomenon. At half-filling any charge modulation is excluded by particle-hole symmetry\cite{charge_transfer} and the local occupation on any layer, including the surface, coincides with the average filling, $n_{\alpha} = 2$. Away from half-filling this is generally not the case. The charge redistribution in the surface region is driven by a narrower local density of states (LDOS) at the surface relative to the bulk LDOS (see Appendix A). However, the resulting electrostatic potential tends to restore the system towards a homogeneous charge distribution. The strength of the electrostatic potential depends on the dielectric constant, for which an appropriate value differs from material to material, and is often not well established. The two panels of \fref{fig:ChargeProfile} show the charge density profile in the surface region for two cases with a rather large charge transfer, i.e., $U/t=1$, for two choices of $t_{12}/t=1.0$ and $t_{12}/t=0.5$, and for three values of $\lambda/t$. The reference case is $\lambda/t =0$, which gives the result without considering the electrostatic potential provided by Poisson's equation, or equivalently, the case with infinite $\epsilon$. The two curves with  $\lambda/t=0.24$ and $\lambda/t=0.9$ correspond to $\epsilon=15$ and $\epsilon=4$, respectively. The dielectric constant of strongly correlated metals is generally taken to be much higher than both these values.\cite{Lunkenheimer10} As it is seen from the solid line in both panels, charge transfer diminishes with increasing distance from the surface and for the third layer, the charge density is essentially the same as the bulk density. Thus, these density changes are very local and we expect that they will be largely unaffected by considering the long-range Coulomb interaction. This intuition is confirmed by the results given by the dashed and dotted lines in both panels. Including the electrostatic potential in the calculation with two typical relatively small values for $\epsilon$ causes only a small suppression of charge transfer. For this reason we will not consider the electrostatic potential in the results that follow.} 

\begin{figure}
\begin{center}
\center{\includegraphics[height=5.0in,width=3.5in,angle=0]{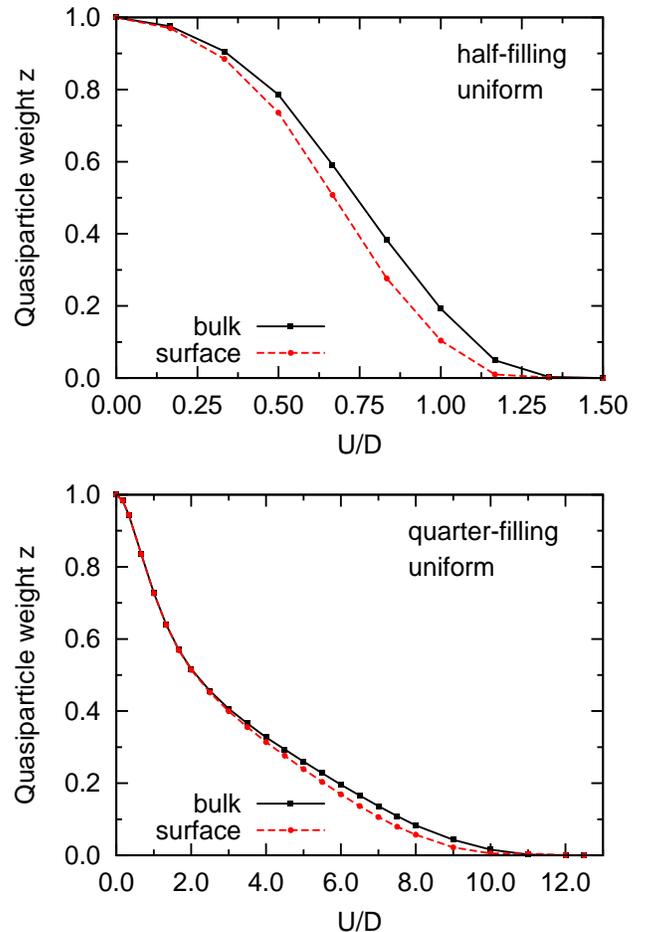}}
\caption{Bulk and surface quasiparticle weights $z$ for the two band Hubbard model as a function of the interaction strength $U/D$. Top panel: half-filling, bottom panel: quarter-filling.}\label{fig:quasiUniform}
\end{center}
\end{figure}

The top panel of \fref{fig:quasiUniform} shows the calculated quasiparticle weight $z_{\alpha=1}=\left( 1-\partial\Sigma_{\alpha=1}(\omega)/\partial\omega\vert_{\omega=0}\right)^{-1}$ of the semi-infinite two-band Hubbard model at $T = 0$ as a function of interaction strength, $U/D$, where $D$ is half the band width. Here,  $\Sigma_{\alpha}(\omega)$ is the self-energy for layer $\alpha$. The quasiparticle weight is a measure of the metallic nature of a system, with $z = 1$  for a non-interacting metal and $z = 0$ for a correlated insulator. As expected, both the bulk and surface quasiparticle weights decrease monotonically as a function of the interaction strength, and they eventually vanish for values of $U$ beyond the critical value, $U_c$. For any value of the interaction strength, the quasiparticle weight of the surface layer $z_{\rm surf}$ is significantly reduced compared to $z_{\rm bulk}$, which can be understood as the effect of the reduced delocalizaton effect of kinetic energy and therefore enhanced effective correlations. The differences between the $z_{\alpha}$ and the bulk $z$ diminish with increasing distance from the surface and for the third layer, the quasiparticle weight is
almost indistinguishable from the bulk $z$ on the scale used (not shown). 

As is seen in the bottom panel of \fref{fig:quasiUniform}, the trend at quarter-filling is different from that at half-filling. In order to understand quasi-particle weight behaviour at quarter-filling we should bring into account charge transfer effects. Indeed, charge depletes from the surface to the bulk in a range of densities including quarter-filling. In other words, while the electron density in the bulk is $n_{\rm bulk}=1.0$, the surface electron density is a little less than quarter-filled, $n_{\rm surf}{{ \atop <} \atop {\sim \atop }}1.0$, which results in reduced correlation effects at the surface. In this case, the characteristics of the surface quasi-particle are approximately the same as those of the bulk quasi-particle, up to a relatively large interaction strength; eventually $z_{\rm surf}$ falls below $z_{\rm bulk}$. 

\begin{figure}
\begin{center}
\center{\includegraphics[height=2.75in,width=3.5in,angle=0]{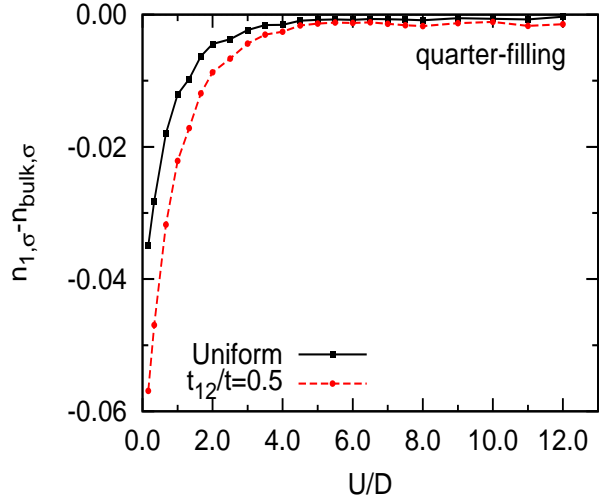}}
\caption{The charge transfer $\Delta n_{\alpha=1}= n_{\alpha=1}-n$ between the surface layer ($\alpha=1$) and the bulk as a function of the interaction strength $U/D$ for uniform $t$ and $t_{12}/t=0.5$. }\label{fig:chargeTransfer}
\end{center}
\end{figure}

A summary of the situation at quarter filling is as follows. Reduced coordination number at the surface leads to two mechanisms: first, an enhanced ratio between the interaction potential and the kinetic energy, and, secondly, charge transfer between the surface and the bulk. The first effect increases correlation effects, while the second decreases them. \fref{fig:chargeTransfer} shows charge transfer as a function of interaction strength. As is evident from this figure, at large interaction strength the surface charge transfer approaches zero and, for these interaction strengths, only the first mechanism remains. As expected for these interaction strengths, enhanced correlation effects lead to a smaller quasi-particle weight at the surface (see the bottom panel of \fref{fig:quasiUniform}).

\begin{figure}
\begin{center}
\center{\includegraphics[height=2.75in,width=3.5in,angle=0]{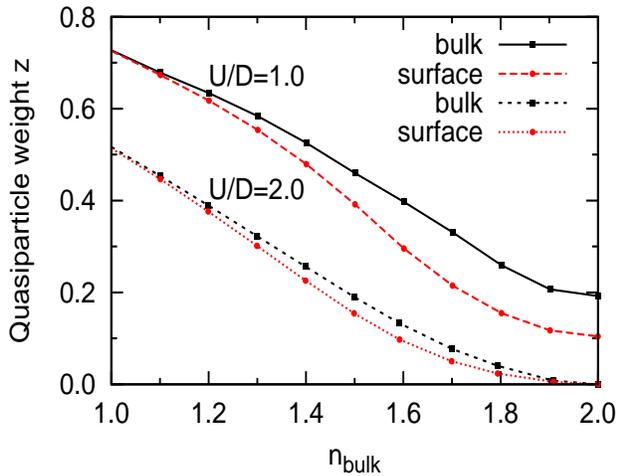}}
\caption{Bulk and surface quasiparticle weights $z$ for the two band Hubbard model as a function of the filling (between quarter and half filling) for two representative interaction strengths $U/D$.}\label{fig:quasiDensity}
\end{center}
\end{figure}

As a function of filling, the variation of quasiparticle weight is shown in \fref{fig:quasiDensity}; at quarter-filling the
surface and bulk values are relatively large and essentially equal to one another. They tend to spread apart as half-filling is approached, although, in cases where $U/D$ is sufficiently large that the material is an insulator, then both surface and bulk values approach one another with value zero. Curves for two representative values of $U/D$ are shown, illustrating this behavior.

\begin{figure}
\begin{center}
\center{\includegraphics[height=5.0in,width=3.5in,angle=0]{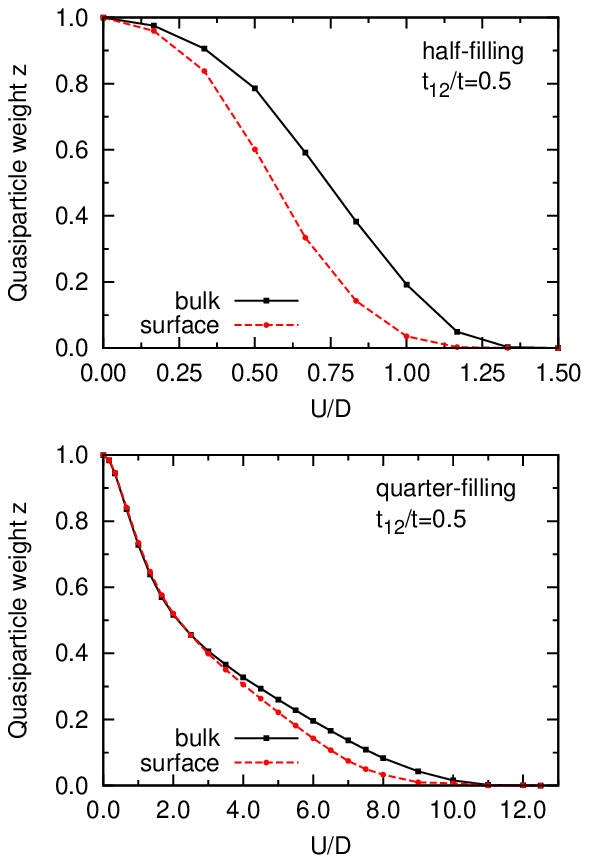}}
\caption{Bulk and surface quasiparticle weights $z$ for the two band Hubbard model as a function of the interaction strength $U/D$. Top panel: half-filling, bottom panel: quarter-filling.}\label{fig:quasiNonUniform}
\end{center}
\end{figure} 

The presence of a surface is likely to alter the effective parameters entering Eq. (\ref{ham}), so we also investigate a case where the hopping between sites on the surface and neighbouring sites immediately below the surface is reduced, 
i.e. $t_{12}/t=0.5$. This choice gives us the opportunity to study the effect of relaxation of the interlayer distance and also changes the competition between the two mechanisms mentioned above. For $t_{12}/t=0.5$ the ratio between the interaction potential and the kinetic energy is further increased. One might expect that the kinetic energy is reduced, but that perhaps the charge transfer may be unaltered, resulting in a change of the balance visible in the bottom panel of \fref{fig:quasiUniform}. However, as is apparent in \fref{fig:chargeTransfer}, a reduced hopping element between the surface and subsurface layers leads to a narrower LDOS at the surface and a larger magnitude of charge transfer. \fref{fig:quasiNonUniform} shows the corresponding calculated quasiparticle weight for the half- and quarter-filled case. While at half-filling the quasiparticle weight is reduced further than in the uniform case (\fref{fig:quasiUniform}), at quarter-filling the two mechanisms continue to balance one another for weak and intermediate coupling strengths; only at very large interaction strengths, where the charge transfer is essentially suppressed (\fref{fig:chargeTransfer}) can we see the effect of further enhanced correlation effects at the surface. For these interaction strengths the $z_{\rm surf}$ approaches zero faster than the uniform case. Actually, at small to intermediate interaction strengths the surface quasiparticle weight is slightly larger than the bulk value of $z$, though it is not evident with the scale used here.

Note that we have confined ourselves to a bipartite lattice, in which there is no charge transfer at half-filling. For a non-bipartite lattice, it is possible to have charge transfer even at half-filling. Therefore, for such lattice structures we expect to have competition between the enhanced ratio of the interaction potential and the kinetic energy, and charge transfer, even at half-filling. 
   
\section{Concluding Remarks}
We have investigated the Mott metal-insulator transition at a solid-vacuum interface at zero temperature in the framework of the semi-infinite two-band Hubbard model at half-filling and quarter-filling. Using the embedding approach to extend dynamical mean-field theory to inhomogeneous systems, it is found that at half-filling, as a result of the surface narrowing of the local density of states, correlation effects are more pronounced at the surface than in the bulk. However, at quarter-filling the surface narrowing of the local density of states leads to an additional effect, namely charge transfer, and the subsequent change in the surface occupation from quarter-filling in fact reduces correlation effects at the surface. It is shown that the interplay of these two mechanisms gives rise to approximately the same quasiparticle weights at the surface and in the bulk. At very large interaction strengths the charge transfer approaches zero and the enhanced correlation causes a reduction of the surface quasiparticle weight.  

\appendix
\section{Charge Transfer Due to a Surface}
\bbf{The band narrowing that occurs at a surface can be understood by referring to moments of the non-interacting local density of states (LDOS)}:
\begin{equation}
M_i^{(m,0)}=\int_{-\infty}^{\infty} E^m \rho_i^{(0)}(E) dE
\end{equation}
\bbf{in which $\rho_i^{(0)}(E)$ is the non-interacting LDOS at site $i$. The variance in the LDOS is given by $\Delta^2 \rho_i^{(0)}=M_i^{(2,0)}-(M_i^{(1,0)})^2 = \sum_{j\ne i} t^2_{ij}$. For a site in the surface $\sum_{j\ne i} t^2_{ij} = 4t^2+t^2_{12}$ while for a site in the bulk it is $\sum_{j\ne i} t^2_{ij} = 6t^2$. The reduced coordination number and probably reduced hopping elements of a site on the surface layer thus implies a reduced width, $\Delta^2 \rho_i^{(0)}$, for the surface LDOS. This key effect of a narrowing of the surface LDOS holds for an interacting system as well.}

\bbf{The problem of how the narrowing of the LDOS leads to charge transfer is straightforward in a one-band model. \cite{Kienert07, Kalkstein71} We will explain first for a non-interacting case and then generalize results to the interacting one. If a completely homogeneous charge distribution is assumed for a moment, the band narrowing implies different Fermi energies for the bulk and the surface. For $n<1$ the surface Fermi energy lies above that of the bulk and for $n>1$ we have the opposite situation and the surface Fermi energy lies below that of the bulk. To restore thermodynamic equilibrium and thus merge the Fermi energies, one has to allow for a charge transfer. Consequently, band narrowing leads to $n_{\alpha=1}<n_{bulk}$ for $n_{bulk}<1$ and $n_{\alpha=1}>n_{bulk}$ for $n_{bulk}>1$.}

\bbf{In the strongly interacting case (where an upper and lower Hubbard band appear), the $0<n<2$ domain of the non-interacting case maps onto the $0<n<1$ domain.}

\begin{acknowledgements}
This work was supported in part by the Natural Sciences and Engineering
Research Council of Canada (NSERC), by ICORE (Alberta), and by the Canadian
Institute for Advanced Research (CIfAR).
\end{acknowledgements}

\bibliographystyle{prsty}

\begin{thebibliography}{10}

\bibitem{Potthoff99_1}
M. Potthoff, and W. Nolting, Phys. Rev. B, \textbf{59}, 2549 (1999).

\bibitem{Potthoff99_2}
M. Potthoff, and W. Nolting, Phys. Rev. B, \textbf{60}, 7834 (1999);
S. Schwieger, M. Potthoff, and W. Nolting, Phys. Rev. B, \textbf{67}, 165408 (2003).

\bibitem{Borghi09} 
G. Borghi, M. Fabrizio, and E. Tosatti, Phys. Rev. Lett. \textbf{102}, 066806 (2009).

\bibitem{Nourafkan09}
R. Nourafkan, M. Capone, and N. Nafari, Phys. Rev. B \textbf{80}, 155130 (2009).

\bibitem{Nourafkan10}
R. Nourafkan, F. Marsiglio, and M. Capone, Phys. Rev. B \textbf{82}, 115127 (2010).

\bibitem{Liebsch03}
A. Liebsch, Phys. Rev. Lett. \textbf{90}, 096401 (2003).

\bibitem{Liebsch06}
H. Ishida, D. Wortmann, and A. Liebsch, Phys. Rev. B \textbf{73}, 245421 (2006).

\bibitem{Maiti98}
K. Maiti, Priya Mahadevan, and D. D. Sarma, Phys. Rev. Lett. \textit{80}, 2885 (1998).

\bibitem{Maiti04}
K. Maiti, Ashwani Kumar, D. D. Sarma, E. Weschke and G. Kaindl, Phys. Rev. B \textbf{70}, 195112 (2004).

\bibitem{Maiti01}
K. Maiti, D. D. Sarma, M. J. Rozenberg, I. H. Inoue, H. Makino, O. Goto, M. Pedio, and R. Cimino, Europhys. Lett. \textbf{55}, 246 (2001).

\bibitem{Sekiyama04}
A. Sekiyama, H. Fujiwara, S. Imada, S. Suga, H. Eisaki, S. I. Uchida, K. Takegahara, H. Harima, Y. Saitoh, I. A. Nekrasov, G. Keller, D. E. Kondakov, A. V. Kozhevnikov, Th. Pruschke, K. Held, D. Vollhardt, and V. I. Anisimov, Phys. Rev. Lett. \textbf{93}, 156402 (2004).

\bibitem{Nourafkan11}
R. Nourafkan and F. Marsiglio, Phys. Rev. B {\bf 83}, 155116 (2011). 

\bibitem{Ohtomo02} A. Ohtomo, D.A. Muller, J.L. Grazul, and H.Y. Hwang, Nature (London) {\bf 419}, 378 (2002).

\bibitem{Okamoto04_1}
S. Okamoto and A. J. Millis, Nature (London) \textbf{428}, 630 (2004).

\bibitem{Okamoto04_2}
S. Okamoto and A. J. Millis, Phys. Rev. B \textbf{70}, 075101 (2004).

\bibitem{Okamoto04_3}
S. Okamoto and A. J. Millis, Phys. Rev. B \textbf{70}, 241104(R) (2004).

\bibitem{Freericks04}
J. K. Freericks, Phys. Rev. B \textbf{70}, 195342 (2004); see also L.
Chen and J. K. Freericks, \textit{ibid}. \textbf{75}, 125114 (2007); J. K. Freericks,
V. Zlatic, and A. M. Shvaika, \textit{ibid}. \textbf{75}, 035133 (2007).

\bibitem{Ishida09}
H. Ishida, and A. Liebsch, Phys. Rev. B \textbf{79}, 045130 (2009).

\bibitem{Georges96}
A. Georges, G. Kotlier, W. Krauth, and M. J. Rozenberg, Rev. Mod.
Phys. \textbf{68}, 13 (1996).

\bibitem{Caffarel94} M. Caffarel and W. Krauth, Phys. Rev. Lett. {\textbf 72}, 1545 (1994).

\bibitem{Koga02}
A. Koga, Y. Imai, and N. Kawakami, Phys. Rev. B \textbf{66}, 165107 (2002). 

\bibitem{Koga07}
K. Inaba and A. Koga, J. Phys. Soc. Jpn. \textbf{76} (2007).

\bibitem{charge_transfer}
For non-bipartite (e.g. fcc) lattice a charge modulation is possible also at half-filling.

\bibitem{Lunkenheimer10}
P. Lunkenheimer, S. Krohns, S. Riegg, S.G. Ebbinghaus, A. Reller and A. Loidl, Eur. Phys. J. Special Topics \textbf{180}, 61–89 (2010).


\bibitem{Kienert07}
J. Kienert and W. Nolting, Phys. Rev. B \textbf{75}, 094401 (2007).

\bibitem{Kalkstein71}
D. Kalkstein and P. Soven, Surf. Sci. \textbf{26}, 85 (1971).


\end{thebibliography}

\end{document}